\documentstyle[multicol,prl,aps,psfig]{revtex}

\newcommand {\kf}{k_{\scriptscriptstyle F}}
\newcommand {\ef}{E_{\scriptscriptstyle F}}
\newcommand {\rd}{\rho_{\scriptscriptstyle D}}
\newcommand {\sdal}{\tilde 
  \sigma_{\scriptscriptstyle D}^{\rm \scriptscriptstyle AL}}
\newcommand {\vf}{v_{\scriptscriptstyle F}}

\begin{document}

\title {Coulomb drag at $\nu=\frac {1}{2}$:
 Composite fermion pairing fluctuations}

\author {Iddo Ussishkin and Ady Stern}

\address {Department of Condensed Matter Physics, The Weizmann
  Institute of Science, Rehovot 76100, Israel}

\date {August 2, 1998}

\maketitle

\begin {abstract}
  We consider the Coulomb drag between two two-dimensional electron
  layers at filling factor $\nu = \frac{1}{2}$ each, using a strong
  coupling approach within the composite fermion picture. Due to an
  attractive interlayer interaction, composite fermions are expected
  to form a paired state below a critical temperature $T_c$.  We find
  that above $T_c$ pairing fluctuations make the longitudinal
  transresistivity $\rd$ increase with decreasing temperature. The
  pairing mechanism we study is very sensitive to density variations
  in the two layers, and to an applied current.  We discuss possible
  relation to an experiment by Lilly {\em et al.}\ [Phys.\ Rev.\
  Lett.\ {\bf 80}, 1714 (1998)].
\end {abstract}

\begin {multicols}{2}

Double layer systems in the quantum Hall regime exhibit a rich variety
of physical phenomena, including transitions between compressible and
incompressible states, as well as interlayer coherence \cite {DP97}.
A powerful probe to such systems is a Coulomb drag experiment, in
which a current $I_1$ is driven through one layer, and the voltage
$V_2$ is measured on the other layer.  The transresistivity is defined
as $\rd = - V_2 / I_1$ for a square sample.  A Coulomb drag
measurement probes the interlayer interaction, the single layer
response, and interlayer correlations.

We concentrate here on the Coulomb drag between identical layers at
filling factor $\nu = \frac {1}{2}$ each, which has been recently
measured by Lilly {\em et al.}\ \cite{LEPW98}.  The weak coupling
limit of this problem was considered using a composite fermion picture
\cite {KM96,Sak97} and an electronic picture \cite {US97}, leading to
three main results.  First, $\rd$ is expected to be much larger
(typically by 3--4 orders of magnitude) than at zero magnetic field,
due to the slow decay of density fluctuations \cite {US97}.  Second,
the low temperature dependence is $\rd \propto T^{4/3}$ \cite
{KM96,Sak97,US97}.  Finally, within a weak coupling theory $\rd$
vanishes at $T = 0$.  The observed low temperature behavior \cite
{LEPW98} seems to contradict the last conclusion, and indicates the
need to go beyond weak coupling theory.

In this paper we consider the Coulomb drag using a strong coupling
approach within the composite fermion picture.  Part of the interlayer
interaction between composite-fermions is attractive \cite {Bon93},
and as a result a composite fermion paired state is expected to form
below a critical temperature $T_c$ \cite {BMN96}.  In electronic
terms, such a state is an incompressible quantum Hall state of the
double layer system, and is expected to have a large transresistivity,
similar to that discussed in Ref.~\cite {VM96}.  Experimentally this
state is not observed, and $\rd$ is always much smaller than the
single layer resistivity $\rho_{xx}$ \cite {LEPW98}.  Motivated by
this experimental finding, we consider here the contribution of
fluctuations of the paired state above $T_c$.  Our central result is
that by including the contribution of pairing fluctuations the
transresistivity becomes
\begin {equation}
  \label {ral} \rd = \frac {\alpha}{4} \rho_{xx}^2 \frac {e^2}{h}
  \frac {T_c}{T - T_c} + \rd^{(0)} \, ,
\end {equation}
where $\alpha \approx 0.12$, and $\rd^{(0)}$ is the weak coupling
contribution \cite {US97}.  This expression is valid provided that $T
- T_c \ll T_c$, but $\rd \ll \rho_{xx}$. Unlike $\rd^{(0)}$, the first
term in (\ref {ral}) {\em increases} with decreasing
temperature. While a similar term affects $\rho_{xx}$ as well, $\rd$
is much more sensitive to it, being typically three orders of
magnitude smaller than $\rho_{xx}$.  Within the mechanism we study,
the drag voltage is always parallel to the current, i.e., there is no
Hall drag.

In the composite-fermion approach \cite {HLR93}, the system of
electrons is transformed by a suitable Chern-Simons transformation to
a system of composite fermions in an average zero field, carrying flux
lines attached to them. In the double layer system, if the interlayer
spacing $d$ is large, a single-layer transformation is to be performed
on each layer separately, leading to the Hamiltonian, in units where
$\hbar = c = 1$,
\begin {eqnarray}
  \label {Hamiltonian} H & = & \frac {1}{2 m_b} \sum_i \int d^2 r \,
 \psi_i^\dagger ({\bf r}) [-i \nabla - \delta {\bf a}_i ({\bf r}) ]^2
 \psi_i ({\bf r}) \nonumber \\ & + & \frac {1}{2} \sum_{ij} \int d^2 r
 \, d^2 r' \, : \delta \rho_i ({\bf r}) v_{ij} ({\bf r} - {\bf r'})
 \delta \rho_j ({\bf r'}) : \, .
\end {eqnarray}
Here, $\psi_i^\dagger ({\bf r})$ is the creation operator for a
composite fermion at point $\bf r$ in layer $i$, $m_b$ is the
electron's band mass, $\delta \rho_i ({\bf r}) = \psi_i^\dagger ({\bf
r}) \psi_i ({\bf r}) - n$ is the density (with the average value $n$
subtracted), $\delta {\bf a}_i ({\bf r})$ is the Chern-Simons gauge
field (with the external vector potential subtracted), and the colons
stand for normal order. The density and the gauge field are related by
$4 \pi \delta \rho_i ({\bf r}) {\bf \hat z} = \nabla \times \delta
{\bf a}_i ({\bf r})$. The Coulomb interaction is given by $v_{ij} (r)
= e^2 / \epsilon \sqrt {r^2 + d^2 (1 - \delta_{ij})}$, where
$\epsilon$ is the bulk dielectric constant.  Since the Chern-Simons
transformation is carried out separately on each layer, the electronic
transresistivity (both longitudinal and Hall components) is identical
to that of the composite fermions.  At $\nu=\frac {1}{2}$, composite
fermions experience no magnetic field on average. Consequently, their
Hall transresistivity vanishes, and so does the electronic one.  We
find this result to hold also around $\nu=\frac {1}{2}$.

The composite fermions interact through the Coulomb interaction and
through an intralayer Chern-Simons interaction introduced by the
transformation (\ref {Hamiltonian}). Within the random phase
approximation, the most singular element of the gauge propagator is
between anti-symmetric transverse components.  This element mediates
an interlayer interaction between transverse currents \cite {Bon93},
\begin{equation}
  \label {u11} U_{11} = \pi \frac {{\bf k}_1 \times {\bf \hat Q}}{m^*}
 \cdot \frac {{\bf k}_2 \times {\bf \hat Q}}{m^*} \frac {Q / \kf}{T_0
 (Q / \kf)^3 -i \Omega} \, .
\end{equation}
Here, composite fermions of momenta ${\bf k}_1$, ${\bf k}_2$ are
scattered to ${\bf k}_1 + {\bf Q}$, ${\bf k}_2 - {\bf Q}$, the energy
transfer is $\Omega$, the Fermi momentum is $\kf = \sqrt {4 \pi n}$,
and the limits $\Omega \ll q \kf / m^*$ and $q \ll \kf$ are used. The
energy scale $T_0$ is, in the limit of strong screening, $T_0 = \pi
e^2 n d / \epsilon$ \cite {US97}. The ratio of $T_0$ and $\ef \sim e^2
\kf / \epsilon$ is proportional to $\kf d$. We assume $T_0 \gg \ef \gg
T$.  We use a dressed interaction vertex, inversely proportional to
the composite fermion effective mass $m^*$. The contribution of all
other components of the gauge propagator is neglected. The strong
singularity of (\ref {u11}) in the limit $Q,\Omega \rightarrow 0$,
which results from the slow dynamics of charge relaxation in the
$\nu=1/2$ state \cite{US97}, poses the main difficulty in the present
calculation.

{From} Eq.~(\ref {u11}) it follows that the interlayer interaction
between fermions moving in opposite directions (${\bf k}_2 = - {\bf
k}_1$) is {\em attractive} \cite {Bon93}.  The only bare interlayer
interaction is the electrostatic one. Thus, this attractive
interaction may be understood in terms of the electronic Coulomb
interaction: a composite fermion in momentum state $\bf k$ carries a
current ${\bf j} \propto \bf k$. In electronic terms, such a current
is accompanied by a chemical potential gradient $(2 h / e^2) {\bf \hat
z}\times{\bf j}$.  The $\nu = \frac {1}{2}$ state is compressible, so
that a chemical potential gradient leads to a density gradient, which
is proportional to ${\bf \hat z} \times {\bf k}$.  The attraction
between a composite fermion of momentum ${\bf k}$ in the first layer
and a composite fermion of momentum ${\bf -k}$ in the second layer
results, then, from the Coulomb attraction between the density
gradients accompanying the two composite fermions, which are
oppositely directed.

The composite fermion attraction is expected to lead to interlayer
Cooper pairing below $T_c$ \cite {BMN96}.  Above $T_c$, pairing
fluctuations enhance the composite-fermion conductivity tensor, and
hence $\rd$. As in conventional superconductivity \cite {ST75}, there
are Aslamazov-Larkin \cite {AL68} and Maki-Thompson \cite {Tho70}
contributions.  Pair breaking mechanisms are significant in this
problem (see \cite{BMN96} and below).  Consequently, the
Aslamazov-Larkin term dominates close to $T_c$.  Physically, it
corresponds to the enhancement of the conductivity due to the
formation of composite fermion Cooper pairs by thermal fluctuations.

We now outline the calculation leading to Eq.~(\ref {ral}). A detailed
presentation will appear elsewhere. We use the finite temperature
Matsubara Green function method. To treat the divergences resulting
from the singular interaction (\ref {u11}) at finite temperatures, we
separate the interaction, depending on the bosonic Matsubara energy
transfer ($\Omega_m = 2 \pi m T$), into a static part, ${\cal U}^{\rm
stat}$, when $\Omega_m = 0$, and a dynamic part, ${\cal U}^{\rm dyn}$,
when $\Omega_m \neq 0$. The infrared divergences result from the
static part, which may be viewed as analogous to a singular disorder
potential that is identical in both layers. These divergences are
exactly canceled in the calculation, a result which is a manifestation
of Anderson's theorem.

We use the following notation: Calligraphic letters (e.g.\ $\cal U$)
denote matrices, with indices corresponding to both momentum and
Matsubara energy. Matrix multiplication is defined with the
accompanying factors,
\begin {equation}
  ({\cal A \ast B})_{{\bf k} m, {\bf k'} m'} = - T \sum_{\tilde m}
  \int \frac {d^2 \tilde k}{(2 \pi)^2} {\cal A}_{{\bf k} m, {\bf
  \tilde k} \tilde m} {\cal B}_{{\bf \tilde k} \tilde m, {\bf k'} m'}
  .
\end {equation}
The trace over a matrix of this type (denoted by ${\rm tr}^*$) and a
multiplication with a vector, which is marked with an arrow, are
similarly defined.

\vspace {0.2cm}
\setlength{\unitlength}{0.48in}
\begin {figure}
  \narrowtext
  \begin{center}
    \begin{picture}(6.5,4)(0.5,-4.875)
      \put(1,-4.875){\psfig{figure=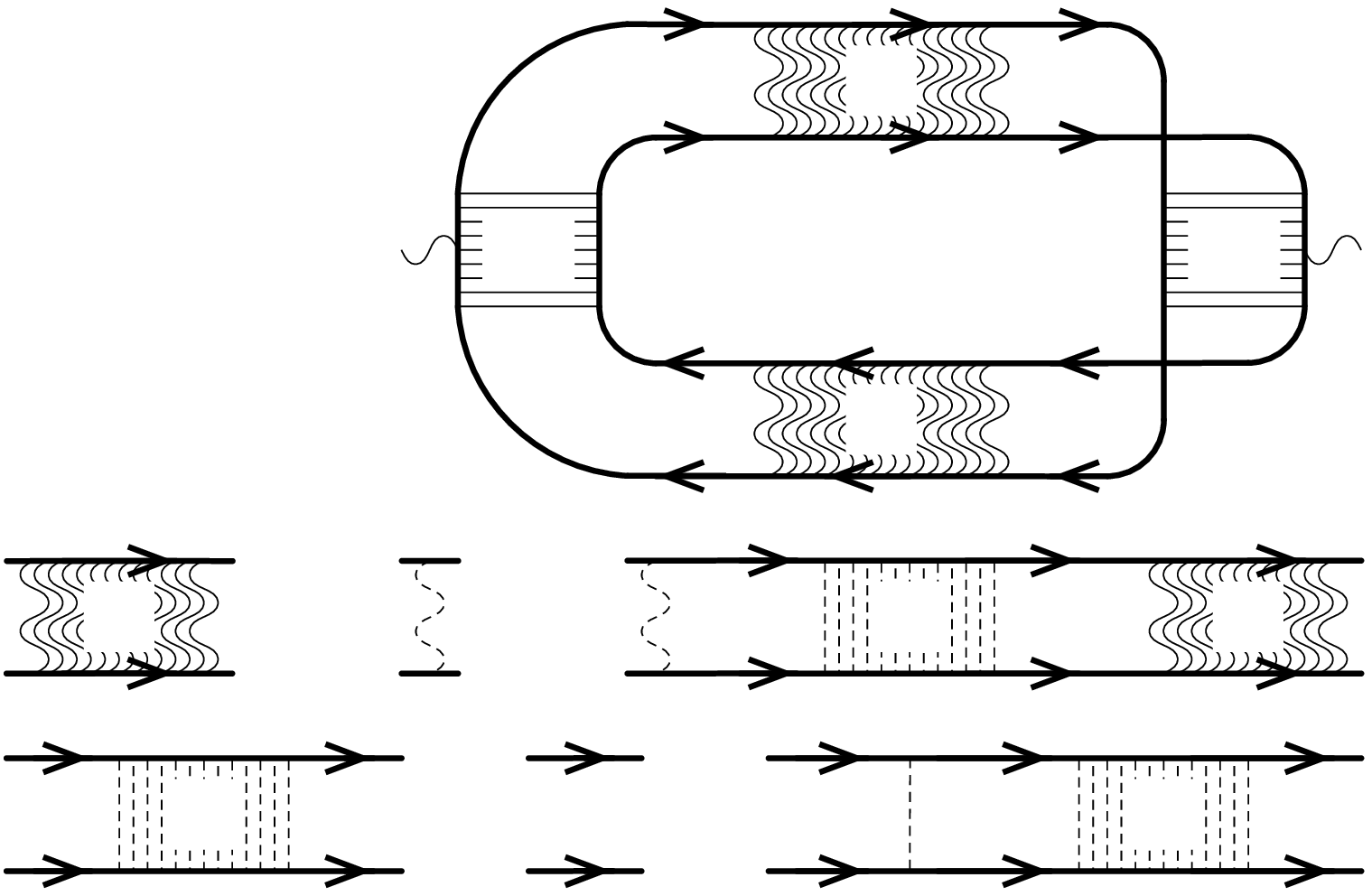,width=6\unitlength}}
      \put(0.5,-2){\makebox(0,0)[l]{$(a)$}}
      \put(0.5,-3.625){\makebox(0,0)[l]{$(b)$}}
      \put(0.5,-4.50){\makebox(0,0)[l]{$(c)$}}
      \put(2.375,-3.625){\makebox(0,0){$=$}}
      \put(3,-4.5){\makebox(0,0){$=$}}
      \put(3.375,-3.625){\makebox(0,0){$+$}}
      \put(4.125,-4.5){\makebox(0,0){$+$}}
      \put(3.775,-1.75){\makebox(0,0)[l]{$\scriptstyle ^\nwarrow - {\bf
        k'}, - \epsilon_{m'}$}}
      \put(3.15,-1.1){\makebox(0,0)[r]{$\scriptstyle {\bf k'} + {\bf q},
        \epsilon_{m'} + \omega_m + \tilde \Omega_{\tilde m
        \searrow}$}}
      \put(3.775,-2.25){\makebox(0,0)[l]{$\scriptstyle _\swarrow - {\bf
        k''}, - \epsilon_{m''}$}}
      \put(3.15,-2.9){\makebox(0,0)[r]{$\scriptstyle {\bf k''} + {\bf
        q}, \epsilon_{m''} + \omega_m \, \!^\nearrow$}}
      \put(4.875,-1.28){\makebox(0,0){$\scriptstyle \cal K$}}
      \put(4.875,-2.76){\makebox(0,0){$\scriptstyle \cal K$}}
      \put(1.52,-3.625){\makebox(0,0){$\scriptstyle \cal K$}}
      \put(6.47,-3.625){\makebox(0,0){$\scriptstyle \cal K$}}
      \put(4.99,-3.625){\makebox(0,0){$\scriptstyle \cal C$}}
      \put(1.89,-4.49){\makebox(0,0){$\scriptstyle \cal C$}}
      \put(6.1,-4.49){\makebox(0,0){$\scriptstyle \cal C$}}
      \put(3.3125,-2.02){\makebox(0,0){$\scriptstyle {\bf q} \cal J$}}
      \put(6.4,-2.02){\makebox(0,0){$\scriptstyle {\bf q} \cal J$}}
    \end{picture}
  \end{center}
\caption {Diagrams for (a) The Aslamazov-Larkin contribution to the
composite-fermion transconductivity $\sdal$ (b) The particle-particle
ladder $\cal K$ (c) The static ladder $\cal C$.  Here, ${\cal U}^{\rm
stat}$ is marked with a dashed line, and ${\cal U}^{\rm dyn}$ with a
dashed wiggly line.  The Green function is denoted by a bold line.
The vertex function ${\bf q} {\cal J}$ in (a) is derived from
Eq.~(\protect \ref {vertex}).}
\label {fig-diagrams}
\end {figure}

The diagrams involved in the calculation of the Aslamazov-Larkin
contribution are given in Fig.~\ref {fig-diagrams}. The
composite-fermion transconductivity is given by
\begin {eqnarray}
  \label {kjkj} \lefteqn {\sdal = \lim_{\tilde \Omega \rightarrow 0}
  \frac {i e^2}{\tilde \Omega} \Biggl\{ -T \sum_m \int \frac {d^2
  q}{(2 \pi)^2} \, q^2 } \\ \nonumber & & \left. \times {\rm tr}^*
  \left[ {\cal J} * {\cal K} ({\bf q}, \omega_m) * {\cal J} * {\cal K}
  ({\bf q}, \omega_m + \tilde \Omega_{\tilde m}) \right] \Biggr\}
  \right|_{i \tilde \Omega_{\tilde m} \rightarrow \tilde \Omega + i
  \delta} .
\end {eqnarray}
Here, ${\cal K} ({\bf q}, \omega_m)$ is the particle-particle ladder
[Fig.~\ref {fig-diagrams}(b) and (c)], with total incoming momentum
$\bf q$ and total incoming (bosonic) Matsubara energy $\omega_m$. The
vertex function is given by ${\bf q} {\cal J}$, where $\cal J$, unlike
$\cal K$, is regular in the long-wavelength low-energy limit at $T =
T_c$. The external frequency is $\tilde \Omega_{\tilde m}$.  The
calculation is carried out to leading order in $T - T_c$.

Because of the separation of the interaction into a static and dynamic
part, the equation for the particle-particle ladder $\cal K$ is also
separated into two parts. The integral equation for $\cal K$ [see
Fig.~\ref {fig-diagrams}(b)] is given by
\begin {equation}
  \label {dyn-ladder-eq} {\cal K} ({\bf q}, \omega_m) = {\cal U}^{\rm
  dyn} ({\bf q}) + {\cal U}^{\rm dyn} ({\bf q}) * {\cal C} ({\bf q},
  \omega_m) * {\cal K} ({\bf q}, \omega_m) \, .
\end {equation}
where the static ladder $\cal C$ is the solution of the integral
equation [see Fig.~\ref {fig-diagrams}(c)]
\begin {equation}
  \label {stat-ladder-eq} {\cal C} ({\bf q}, \omega_m) = {\cal G}
 ({\bf q}, \omega_m) + {\cal G} ({\bf q}, \omega_m) * {\cal U}^{\rm
 stat} ({\bf q}) * {\cal C} ({\bf q}, \omega_m) \, .
\end {equation}
Here,
\begin {eqnarray}
  {\cal G}_{{\bf k} m, {\bf k'} m'} ({\bf q}, \omega_{\tilde m}) & = &
  - \frac {(2 \pi)^2}{T} \delta_{mm'} \delta^2 ({\bf k} - {\bf k'}) \\
  \nonumber & \times & G ({\bf k} + {\bf q}, \epsilon_{m} +
  \omega_{\tilde m}) G (-{\bf k}, - \epsilon_{m}) \, ,
\end {eqnarray}
where $\epsilon_m = (2 m + 1) \pi T$ is a fermionic Matsubara
energy. The static ladder is analogous to the Cooperon in disordered
systems. For the Green function, $G (k, \epsilon_m) = [i \epsilon_m -
\epsilon_k + \Sigma (k, \epsilon_m)]^{-1}$, we use the self-consistent
Born approximation for the static interaction,
\begin {equation}
  \label {sigma} \Sigma (k, \epsilon_m) = T \int \frac {d^2 k'}{(2
  \pi)^2} \, {\cal U}^{\rm stat}_{{\bf k} m, {\bf k'} m} ({\bf q} = 0)
  G (k', \epsilon_m) \, .
\end {equation}
Eq.~(\ref {sigma}) results from the observation that the most singular
term in the intralayer interaction is $-U_{11}$. The factor $T$
appeared from the separation of the Matsubara sum. The contribution of
${\cal U}^{\rm dyn}$ to the self-energy is non-divergent.  Thus, its
inclusion here is not essential. The angular averaging of ${\cal
U}^{\rm stat}$ is, for small $|k-k'|$,
\begin{equation}
  \label {ustat} \int \frac {d \theta}{2 \pi} \, {\cal U}^{\rm
  stat}_{{\bf k} m, {\bf k'} m'} ({\bf q} = 0) = - \frac {\pi}{4}
  \frac {\kf^3}{m^{*2} T_0 |k - k'|} \delta_{mm'} \, ,
\end{equation}
where $k$ and $k'$ are assumed be close to $\kf$, and $\theta$ is the
angle between $\bf k$ and $\bf k'$. The divergence due to the
interaction is thus logarithmic. To regularize the divergence, the
singular term $1 / |k - k'|$ may be replaced with $\tanh [\Lambda (k -
k')] / (k - k')$, and the limit $\Lambda \rightarrow \infty$ taken at
the end of the calculation. In our case the terms involving ${\cal
U}^{\rm stat}$ are canceled exactly.

We first describe the solution of the particle-particle ladder
[Eqs.~(\ref {dyn-ladder-eq})--(\ref{sigma})] at ${\bf q} = \omega_m =
0$, where the pairing instability emerges at the critical temperature
$T_c$. The static ladder $\cal C$ has a singularity at $\epsilon = 0$
(after analytic continuation $i \epsilon_m \rightarrow \epsilon + i
\delta$), where resonant scattering occurs. The terms contributing to
the singular part ${\cal C}^s$ are those with multiple scattering
lines.  As a result, ${\cal C}^s$ is separable in the momentum
variables:
\begin {equation}
 {\cal C}^s_{{\bf k} m, {\bf k'} m'} = - \frac {\delta_{mm'}}{T} \frac
  {1}{\pi N_0} \frac {{\rm Im} G^+ (k,0) \, {\rm Im} G^+
  (k',0)}{|\epsilon_m|} \, ,
\end {equation}
where $N_0$ is the density of states at the Fermi energy and the
energy is measured from the Fermi surface.

The particle-particle ladder ${\cal K} ({\bf q} = \omega_m = 0)$ is
singular at $T = T_c$. To calculate the singular part ${\cal K}^s$
[using Eq.~(\ref {dyn-ladder-eq})], an angular averaging over the
Fermi surface of ${\cal U}^{\rm dyn} ({\bf q} = 0)$ is needed [cf.\
(\ref {ustat})]. For $m \neq m'$ it is
\begin{equation}
  \bar U^{\rm dyn}_{m,m'} = -\frac {2 \pi}{3 \sqrt {3}} \frac
  {\kf^2}{m^{*2} T_0^{2/3} |\Omega_{m-m'}|^{1/3}} \, .
\end{equation}
This result assumes $T_0 (|k - k'| / \kf)^3 \ll |\Omega_m|$, an
assumption justified a posteriori by Eq.~(\ref {tc}). After averaging,
the dynamic interaction is momentum independent, and thus ${\cal K}^s$
is momentum independent and separable in the energy variables. We
therefore write
\begin {equation}
 \label {ks} {\cal K}^s_{{\bf k} m, {\bf k'} m'} = - \frac {1}{N_0} \,
 \frac {T_c}{T - T_c} \kappa_m \, \kappa_{m'} \, .
\end {equation}
Using Eq.~(\ref {dyn-ladder-eq}), at $T=T_c$ the vector $\roarrow
\kappa$, whose ${\bf k} m$ component is $\kappa_m$, satisfies
$\roarrow \kappa = {\cal U}^{\rm dyn} * {\cal C} * \roarrow \kappa$,
i.e.,
\begin {equation}
  \label {eigen-eq} \kappa_m = - T \sum_{m' \neq m} \bar U_{m,
  m'}^{\rm dyn} \, \frac {\pi N_0}{|\epsilon_{m'}|} \, \kappa_{m'} \,
  .
\end {equation}
The highest temperature for which (\ref {eigen-eq}) has a solution is
$T_c$. Note that Eq.~(\ref {eigen-eq}) is independent of ${\cal
U}^{\rm stat}$. Solving (\ref {eigen-eq}) numerically, we find
\begin {equation}
  \label {tc} T_c \approx 0.32 \frac {\ef^3}{T_0^2} \, ,
\end {equation}
where we used $N_0 = m^* / 2 \pi$ and $\ef = \kf^2 / 2 m^*$.
Eq.~(\ref{tc}) is in agreement with Ref.~\cite {BMN96}. As pointed out
by Bonesteel {\em et al.}\ \cite {BMN96}, $T_c$ is reduced by
symmetric density fluctuations, which the composite fermions view as a
fluctuating pair-breaking magnetic field.  Impurities, which exert a
different disordered potential in each of the two layers, further
suppress $T_c$.  Additional controllable mechanisms for $T_c$
suppression are discussed below.

An expansion of the results for small $\bf q$ and $\omega$ is needed
for the denominator of ${\cal K}^s$ in (\ref {ks}), while the
numerator is regular in the long-wavelength low-energy limit at $T =
T_c$.  At finite $\bf q$ and $\omega_m > 0$, Eq.~(\ref {eigen-eq}) is
written with $\bar U^{\rm dyn}$ and the denominator of ${\cal C}^s$
taken at $\bf q$ and $\omega_m$, and the sum limited to $m' \geq 0$
and $m' < -m$.  The temperature at which Eq.~(\ref {eigen-eq}) has a
solution, which is $T_c$ at ${\bf q} = \omega_m = 0$, is then expanded
for small $\bf q$ and $\omega$ (after analytic continuation $i
\omega_m \rightarrow \omega + i \delta$). We find,
\begin {equation}
  \label {kqw} {\cal K}_{{\bf k} m, {\bf k'} m'}^{s} = - \frac
 {1}{N_0} \, \frac {T_c}{T - T_c - i \alpha \omega + \eta q^2}
 \kappa_m \, \kappa_{m'} \, .
\end {equation}
Numerically we find $\alpha \approx 0.12$. The calculation of $\eta$,
and of the vertex function is difficult since both depend on ${\cal
U}^{\rm stat}$. However, they are related by
\begin {equation}
  \label {etaj} \roarrow \kappa^\dagger * {\cal J} * \roarrow \kappa =
  \frac {2 N_0 \eta}{T_c} \, .
\end {equation}
Here, the expansion of the vertex function to first order in ${\bf q}$
at $\omega_m = \tilde \Omega_{\tilde m} = 0$ is ${\bf q} {\cal J}$.
This equality is a consequence of a Ward identity.  It is shown by
writing $\cal J$ as
\begin {eqnarray}
  \label {vertex} {\cal J} & = & \frac {1}{2} \sum_{\mu = x,y} \Bigl
  ( \partial^2_{q_\mu} {\cal C}^s + 2 \partial_{q_\mu} {\cal C}^s *
  {\cal U}^{\rm dyn} * \partial_{q_\mu} {\cal C}^s \\ \nonumber & + &
  2 \partial_{q_\mu} {\cal C}^s * \partial_{q_\mu} {\cal U}^{\rm dyn}
  * {\cal C}^s + 2 {\cal C}^s * \partial_{q_\mu} {\cal U}^{\rm dyn} *
  \partial_{q_\mu} {\cal C}^s \Bigr) \, ,
\end {eqnarray}
and comparing with the expansion for $\eta$, described above.  We note
that diagrams with $\psi^\dagger a^2 \psi$ vertices appear in (\ref
{vertex}) when a derivative with respect to $q_\mu$ is applied to a
transverse current vertex of $\cal U$ [see Eq.~(\ref {u11})].
Eq.~(\ref {etaj}) leads to a cancelation of both $\cal J$ and $\eta$
from $\rd$.

Finally, $\rd$ may be calculated. The calculation of the
composite-fermion transconductivity $\sdal$ (\ref {kjkj}), using
Eqs.~(\ref {kqw}) and (\ref {etaj}), is similar to the
Aslamazov-Larkin calculation \cite {AL68}. The conductivity tensor is
then inverted to obtain the composite-fermion transresistivity, which
is equal to the (physical) electronic transresistivity $\rd$. Assuming
$\rd \ll \rho_{xx}$, we finally obtain Eq.~(\ref {ral}). We now
comment on some aspects of that result.

First, we note that {\em pairing fluctuations lead to an increase of
 the transresistivity as the temperature is decreased,} in sharp
 contrast to the temperature dependence of the weak coupling
 contribution.

Second, the pairing contribution to $\rho_D$ is much more sensitive to
a small density difference $\Delta n$ between the layers than the weak
coupling one.  Such a density difference amounts to a difference
$\Delta \mu = (\partial \mu / \partial n) \Delta n$ between the
chemical potentials of the two layers, where $\partial \mu / \partial
n \sim e^2 / \epsilon \kf$. Pairing between the layers is gradually
suppressed by $\Delta\mu$, in a way similar to the suppression of
conventional Cooper pairing by spin polarization. Complete suppression
of the pairing contribution to $\rd$ happens when $\Delta\mu\approx
T_c$, i.e., when $\Delta n / n \approx T_c / \ef$.  Similarly, the
pairing mechanism is also sensitive to a symmetric deviation of the
density in both layers to a filling factor $\nu = \frac {1}{2} +
\Delta \nu$. In that case both layers are subject to a uniform
Chern-Simons magnetic field $\Delta B\propto \Delta\nu$, which
suppresses pairing. In realistic samples we expect both dependences,
on symmetric and asymmetric density deviation from $\nu=1/2$, to be
mostly determined by the inevitable disorder-induced fluctuations in
the density.

Third, we note that the pairing contribution is also easily suppressed
by the current ${\bf j}_1$ flowing in one of the layers.
Qualitatively, such a current shifts the composite fermion Fermi
sphere in the current carrying layer relative to the one in the other
layer by $\Delta{\bf k}\propto {\bf j}_1$, and thus suppresses the
critical temperature to $T_c-\eta(\Delta k)^2$. In a very naive
estimate $\eta \sim \vf^2 / T_c$ and $\Delta k \sim m^* j_1 / n e$.

Experimentally, Lilly {\em et al.}\ \cite {LEPW98} observed a
non-monotonic temperature dependence of $\rd$ over some narrow ranges
of magnetic field in the vicinity of $\nu = \frac {1}{2}$. The precise
position of these ranges seems to be sample specific, and the
non-monotonic temperature dependence is observed only at very low
currents.  The analysis above suggests that this phenomenon is the
result of {\em pairing fluctuations in those parts of the sample at
which both layers are at filling factor very close to $\frac {1}{2}$.}
The precise values of magnetic field at which such parts exist depend
on the specific disorder configuration.

A quantitative comparison between theory and experiment is hard to
achieve at this stage, since the one parameter in Eq. (\ref{ral}),
$T_c$, is hard to calculate, and there is not enough experimental data
to allow for a fit.

It was recently suggested \cite {Yan98} that the low temperature
behavior of $\rd$ observed by Lilly {\em et al.}\ was due to the
interlayer distance being close to the critical value where the two
layers form an incompressible state of the $(1,1,1)$ type.  The
mechanism we suggest is different from the one leading to a $(1,1,1)$
state. In particular, {\em the Hall transresistivity is zero in our
mechanism}, in contrast to the finite value predicted in \cite{Yan98}.
A measurement which separates the longitudinal and Hall components of
the transresistivity may distinguish between the two mechanisms.

In summary, we have calculated the contribution of pairing
fluctuations to the transresistivity in a system of two layers at
$\nu=\frac {1}{2}$ each. We find that above the critical temperature,
pairing fluctuations, introduced by the interlayer attractive
interaction between composite-fermions, enhance the
transresistivity. In particular, the transresistivity does not vanish
as the low temperature limit is approached.

We thank J.~P. Eisenstein, B.~I. Halperin, A. Kamenev,
D.~E. Khmelnitskii, M.~P. Lilly, and A.~H. MacDonald for stimulating
discussions. This research was supported by the U.S.-Israel Binational
Science Foundation (95/250) and the Minerva Foundation (Munich,
Germany).

Note added: A related preprint (F. Zhou and Y.~B. Kim,
cond-mat/9807321), with which some of our results disagree, has been
very recently posted on the electronic archive.

\end {multicols}

\end {document}